\documentclass[apjl]{emulateapj}
\usepackage[utf8]{inputenc}
\usepackage{comment}
\usepackage{ifthen}
\usepackage{amsmath}
\usepackage{amsfonts}
\usepackage{amssymb}
\usepackage{multirow}
\usepackage{subfigure}
\usepackage{epsfig}
\usepackage[normalem]{ulem}
\usepackage{color}

\newcommand{\rms}{\textrm{RMSD}}

\newcommand{\ele}{e^{-}}

\newcommand{\eskpp}{\ele_{sky}}
\newcommand{\estar}{\ele_{\star tot}}

\def\lsim{\raise0.3ex\hbox{$<$}\kern-0.75em{\lower0.65ex\hbox{$\sim$}}}
\def\gsim{\raise0.3ex\hbox{$>$}\kern-0.75em{\lower0.65ex\hbox{$\sim$}}}
\begin{document}

\title{Observations of the M82 SN with the Kilodegree
Extremely Little Telescope}

\author{Robert J. Siverd\altaffilmark{1}, Ariel Goobar\altaffilmark{2}, 
Keivan G.\ Stassun\altaffilmark{1,3}, Joshua Pepper\altaffilmark{4,1}}
\altaffiltext{1}{Vanderbilt University, Deptartment of Physics \& Astronomy,
VU Station B 1807, Nashville, TN 37235, USA}
\altaffiltext{2}{The Oskar Klein Centre, Department of Physics, Stockholm University,
    SE 106 91 Stockholm, Sweden}
\altaffiltext{3}{Fisk University, Physics Department, 1000 17th Ave.~N., Nashville, TN 37208, USA}
\altaffiltext{4}{Lehigh University, Department of Physics, 413 Deming Lewis Lab, 16 Memorial Drive East
Bethlehem, PA  18015, USA}

\begin{abstract}
We report observations of the bright M82 supernova 2014J serendipitously
obtained with the Kilodegree Extremely Little Telescope (KELT).
%short-timescale light variations and provides
%a stringent constraint on the size of the progenitor.
The SN was
observed at high cadence for over 100 days, from pre-explosion, to early rise and peak times,
through the secondary bump.
The high cadence KELT data with high S/N is 
completely unique for SN 2014J and for any other SNIa, with the exception of the (yet)
unpublished Kepler data. Here, we report determinations of 
the SN explosion time and peak time. 
%and reddening as constrained by the secondary bump. 
We also report measures of the ``smoothness" of the
light curve on timescales of minutes/hours never before probed, and we use this to
place limits on energy produced from short-lived isotopes or inhomogeneities in the explosion or
the circumstellar medium. From the non-observation of significant perturbations of the
light curves, we derive a $3\sigma$ upper-limit corresponding to $8.7 \times 10^{36}$ erg \,s$^{-1}$
for any such extra sources of luminosity at optical wavelengths.
%that have never been sought for. 
%As a by-product we should accurately
%measure the time of maximum and the shape of the secondary bump. The latter 
%could become quite interesting since its explanation remains challenging to
%modelers, e.g., we see now that there is no such bump at mid-IR wavelengths,
%contrary to the models available. Again, the high cadence measurements around 
%the secondary peak add something new (and possibly unexpected).
%We approach this data from a purely empirical point of view, and set limits on
%possible lightcurve features at all available epochs.
\end{abstract}

\section{Introduction\label{sec:intro}}

The study of supernovae to date has been mostly limited to cadences of
days or longer. Exploring shorter time scales could shed new light
on the nature of the progenitor system, the physics of the
explosions, and possibly also the circumstellar environment. In the
first study of its kind, \citet{2011Natur.480..344N,
  2012ApJ...744L..17B} used observations from the first hours of the
optical onset of supernova (SN) SN\,2011fe in M\,101 to confirm the
theoretical expectation that Type Ia supernovae (SNe~Ia) are the
result of thermonuclear explosions of compact objects, C/O white
dwarfs. There is considerable debate concerning the possible systems
leading to SNe~Ia, in particular the core set (``normals'') used
as distance indicators in cosmology. A well-sampled lightcurve at early
times could yield evidence of interaction with a donor star
\citep[e.g.,][]{Kasen2010}, as
expected from the single degenerate (SD) model
\citep{1973ApJ...186.1007W}. The double degenerate model (DD), where
two white dwarfs merge
\citep{1981NInfo..49....3T,1984ApJS...54..335I,1984ApJ...277..355W},
could also leave a signature in the form of an accretion disk, lasting
only for a very short period of time, not sampled by current
observational surveys. Highly cadenced observations can also be used
to explore the production of short-lived radioactive isotopes in the
thermonuclear explosion, as well as the interaction between the ejecta
and thin shells of circumstellar material surrounding the exploding
star. SN\,2014J, the closest ``normal'' SNIa  since the beginning of
modern CCD astronomy  \citep{2014ApJ...784L..12G,2014ApJ...788L..21A,
 2014ApJ...790....3K,2014arXiv1405.3970M,2014arXiv1405.3677F,2014arXiv1408.2381B}, is a particularly well-suited object for 
high-precision studies that could help advance our understanding
of these objects that are so crucial for cosmology.

\section{Data and Methods\label{sec:data}}
%{\bf [Rob, describe KELT data and a little bit about KELT itself.]}
The M82 SN 2014J was serendipitously observed by the Kilodegree Extremely
Little Telescope North (KELT-N) as part of its routine monitoring of the
northern sky. The KELT-N telescope is a robotic telescope designed to search 
for transiting extrasolar planets around bright stars. The optical system 
consists of an Apogee AP16E CCD ($4096 \times 4096$ $9\mu$m pixels)
illuminated by a Mamiya wide-field, medium-format camera lens with 80mm 
focal length and 42mm aperture ($f$/1.9). The resulting images subtend
$26^{\circ} \times 26^{\circ}$ at about $23"$ pixel$^{-1}$. It
employs a Kodak Wratten \#8 red-pass filter in front of the lens to mitigate
the photometric effects of atmospheric reddening (which are most severe at
blue wavelengths). The resulting bandpass resembles a widened Johnson-Cousins
$R$ band with effective wavelength $\lambda_{\rm eff} \approx 691$ nm and width
$\approx 318$ nm. This system is mounted on a Paramount ME robotic telescope mount.
KELT-N typically achieves $\sim$1\% r.m.s.\ photometric precision for $V\approx$ 8--10, 
comparable to the brightness of M82 and of 
SN 2014J at peak brightness. The telescope hardware and operations are detailed more
thoroughly in \citet{KN_telescope} and \citet{2012ApJ...761..123S}. 

Since SN 2014J sits on top of a bright and highly spatially variable backround due to the underlying host galaxy, it was necessary to adjust our standard data reduction procedures which have been optimized for bright individual stars in the Milky Way.
In this section we describe the data that were obtained with
KELT-N, as well as the modifications to the standard calibration and data processing that were required for this object. We emphasize that the light curve of SN 2014J presented here is not strongly dependent on the specific choice of various data reduction parameters discussed below. Rather, our intent here is to fully document the detailed procedures that were required to extract a high quality light curve of SN 2014J from a telescope system that was designed and optimized for a very different type of object.

The final light curve that we present and analyze below is provided in Table~\ref{tab:data}.

\begin{deluxetable}{rrr}
\tabletypesize{\footnotesize}
\tablewidth{\linewidth}
\tablecolumns{3}
\tablecaption{KELT light curve of SN 2014J\label{tab:data}}
\tablehead{
    \colhead{JD (TT)} & 
    \colhead{Flux (ADU)} & 
    \colhead{Error (ADU)}
}
\startdata
2456607.033097  &  -64.77650  &  171.68029\\
2456609.038233  &  127.30780  &  198.71798\\
2456611.033137  & -207.46240  &  151.60861\\
2456616.011700  &  -55.97750  &  209.25459\\
2456616.026432  &  -80.47890  &  209.27662\\
\enddata
\tablecomments{The full table is provided in the electronic journal. A portion is shown here for guidance regarding form and content.}
\end{deluxetable}

\subsection{Data}

KELT-N began observing new fields near the North Celestial Pole (NCP)
in the fall of 2013. One of those fields by chance includes the bright galaxy
M82. As part of its routine robotic observing operations, KELT-N observes the
field containing M82 several times per night, on average, although the specific
cadence varies from night to night depending on Moon and observing conditions.
In particular, 
SN 2014J exploded near full moon. Due to KELT's normal strategy of avoiding observing fields near the full moon,
we fortuitously obtained a larger-than-normal number of observations of the M82 field right around the time of explosion.
Increased photometric noise in the KELT images of the M82 field due to increased sky brightness from the moon scatter was thus offset by an increased number of data points.

The KELT-North telescope uses a German Equatorial mount, causing a 180-degree rotation between images acquired on the East and West sides of the meridian. 
In addition, due to a tilt in the optical system, the KELT-North PSF variations are not circularly
symmetric. It is therefore possible (and common) to see very different
PSFs between East and West. At the location of M82, the effective West 
PSF size is $\sim1/3$ smaller than its East counterpart. The
smaller effective PSF area admits less sky flux and therefore has
better precision, particularly among fainter sources that are dominated 
by sky flux.
As a result, observations in each orientation must be reduced independently. In order to avoid systematics associated with stitching together data from the two telescope orientations, in our analysis below we generally prefer to utilize the data from only the West orientation because in general the photometry is of higher quality. 
However we do incorporate the East orientation data as well for increased precision in our light curve feature timing measurements (see below).

In total, we have 1869 science-grade images (980 east, 889 west) 
%of KELT-N survey field 26 
acquired between 08 October 2013 and 14 June 2014 (JD 2456573.963 to 2456822.692).
The 980 east images were acquired between 08 October 2013 and 01 April 2014 
(JD 2456573.963 to 2456748.622) and the west images acquired between 10 November 2013 and 14 June 2014 (JD 2456607.033 to 2456822.692).
The combined data set spans the nominal explosion time of JD 2456672.25 (UT) \citep[see][and see also Sec.~\ref{sec:t0}]{2014ApJ...783L..24Z},
the nominal peak time of JD 2456690.75 (UT), 
and well into the late-stage dimming of the event.
%{\bf [Wordy, rephrase.]}

% East data span: 08 October 2013 and 01 April 2014 (JD 2456573.963--2456748.622)
% West data span: 10 November 2013 and 14 June 2014 (JD 2456607.033--2456822.692)

\subsection{Dark and Flat Calibration}
Nearly half of KELT-N dark frames during the season when SN 2014J was observed exhibit oscillating electrical noise with
non-negligible amplitude, and with a length scale a few times larger than a typical
point-spread-function full-width-at-half-maximum. 
This electrical noise arises from electronics that are activated only during dark frame acquisition and therefore does not affect the science frames.
If the dark frames are stacked without correction, the resulting master dark frame exhibits pattern noise. In
combination with pointing drift, these patterns induce correlated noise in
light curves that is difficult to correct, particularly in a case such as SN 2014J where we hope to recover the pre- and post-explosion flux to the lowest levels possible. 

Therefore, 
we generated a new, low-noise master dark frame using images from a previous season and used
it to correct all images in this study. 
%All available darks with appropriate CCD temperature were used. 
In each individual dark frame, we identified and removed the electrical noise in Fourier space. The resulting clean darks were level-matched to compensate for varying
bias and then median-stacked with per-pixel outlier rejection.

We flat-field the dark-subtracted science images with our standard master flat
\citep[see][for details of construction of the master flat]{2012ApJ...761..123S}.
%{ \bf should we include more details about making the flat, or point to the specific section of the KELT-1 paper?}

\subsection{Gradient Correction and Cloud Removal}
After dark and flat calibration, we remove complex background spatial variations 
%arising from the bright and spatially variable host galaxy 
with a two-step process. Doing so before the image subtraction procedure prevents corruption of the convolution kernel and improves results, particularly with poor weather and high airmass (see below).

As the first stage of background removal, we remove the overall sky brightness 
gradient on the largest spatial scales. To do this, we fit a second-degree polynomial 
to each image with a Huber M-estimator \citep{huber_book}, a robust regression procedure
that automatically
ignores contaminated (non-sky) pixels. We then subtract the best-fit polynomial from 
each image to remove large-scale sky gradients. Afterward, we add a constant such that 
the original median image value is restored.

After removing the large-scale gradient polynomial, we identify and mask extreme outliers (above 99th percentile) which include stars, passing airplanes, etc. We then apply a 201x201-pixel 45th percentile smooth to the masked image to map out sky variations on smaller spatial scales in a non-parametric way. We finally subtract the polynomial fit and smoothed image from the calibrated image and add a constant so that again the median image value is preserved. 

\subsection{Image Subtraction}
The heart of the KELT light curve production process is image subtraction-based aperture photometry.
In our standard procedure,
KELT-N images are divided into 5x5 subframes of 816x816 pixels, each of which is processed separately. Unfortunately, by chance, M82 falls on an interior corner of the usual KELT-N subframes. 
This led to poor photometric results from our standard pipeline 
\citep[see][for further details]{2012ApJ...761..123S}. In order to improve the quality 
of the photometric extraction for SN 2014J, we chose to proceed with a single 816x816 
subframe centered on M82, extracted after image registration. 
Then we proceeded to process this subframe via our usual image subtraction procedures 
with usual parameter choices. Importantly, we build our
reference image exclusively from pre-explosion data to simplify the 
analysis as much as possible. Finally, we obtain an accurate WCS coordinate
solution for our reference image using Astrometry.net \citep{2010AJ....139.1782L}.

In addition to the supernova, we extract light curves for numerous other
stars in the area and use these to verify pipeline performance and estimate
accurate photometric uncertainties. Including SN 2014J, we extracted 1874 Western
and 1699 Eastern light curves. 
Because the supernova is not visible
in our reference image, we adopt the R.A. and Dec coordinates from 
\citet{2014ApJ...784L..12G} and convert these to pixel coordinates using the 
aforementioned reference image WCS.

\subsection{Accurate Photometric Uncertainties}
\begin{figure}
\centering
\includegraphics[width=0.7\textwidth]{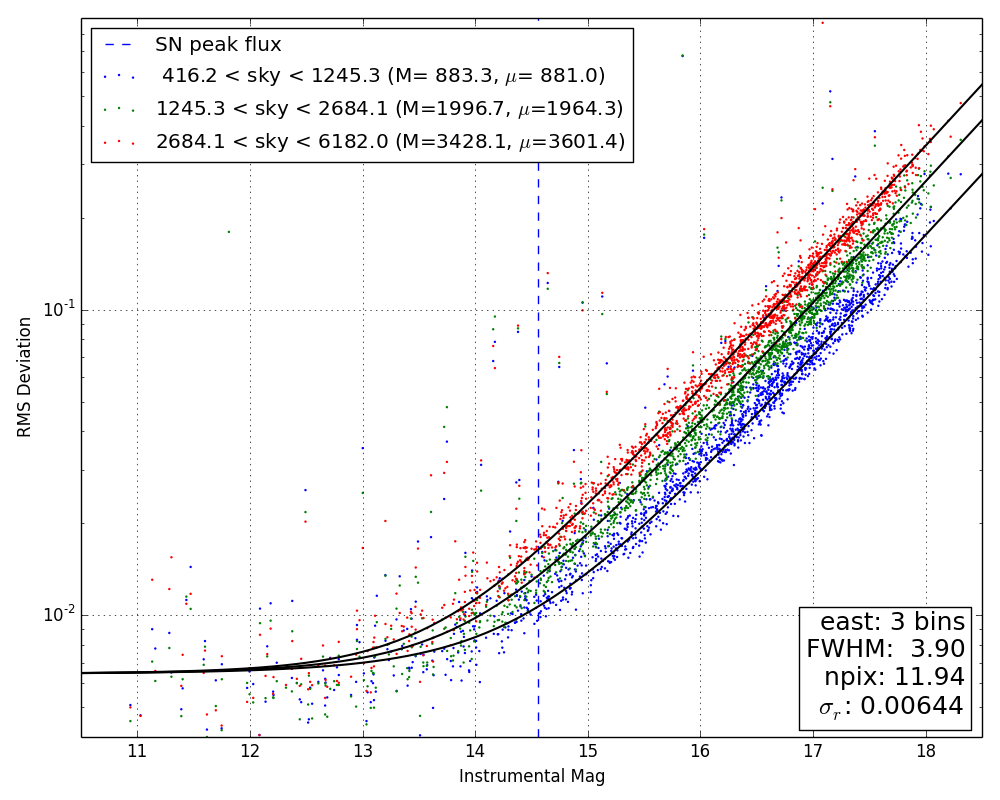}
\includegraphics[width=0.7\textwidth]{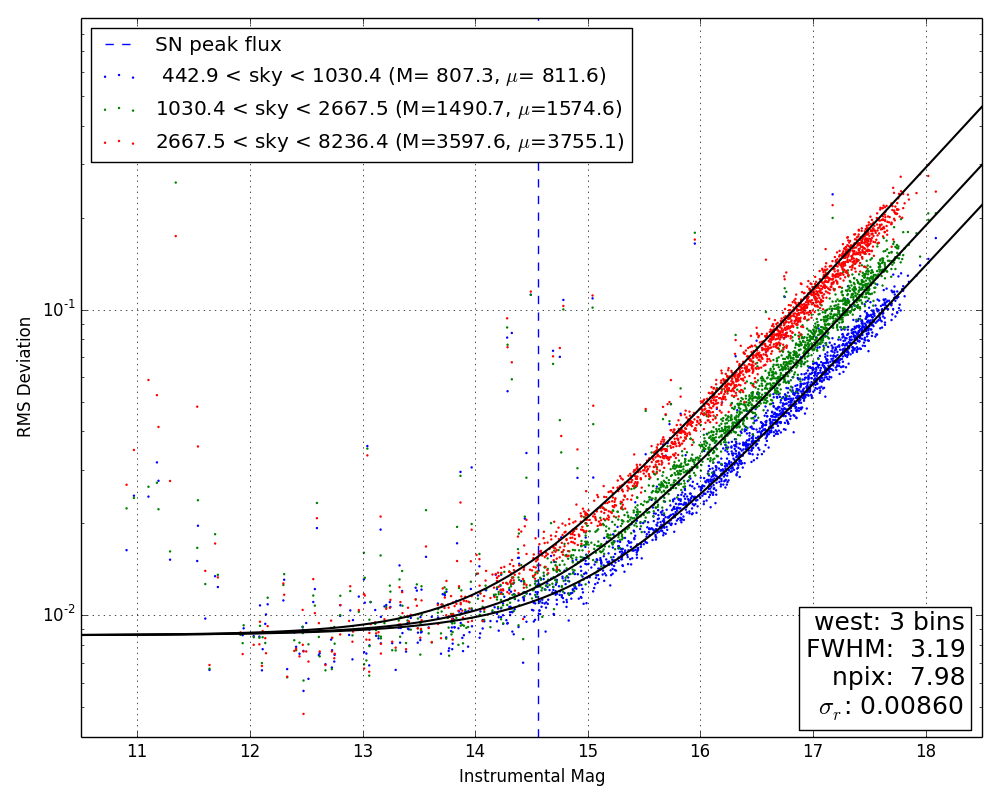}
\caption{KELT noise model for east data (left) and west data (right).
We characterize
our photometric uncertainties empirically using an ensemble of nearly
2000 stars in the vicinity of M82. We find that the KELT-North photometric
performance (i.e., light curve scatter as a function of source brightness
and sky level) is well approximated by a Poisson noise aperture photometry
model with a constant $\sigma_r$ noise floor. Once the aperture
area and noise floor are known, reliable photometric uncertainties may be 
calculated directly for any target.
\label{fig:rms_model}}
\end{figure}

Due to optical vignetting and our preprocessing routines, the standard 
photometric uncertainties produced by our ISIS-based 
\citep{1998ApJ...503..325A,2000A&AS..144..363A} pipeline are often not reliable.
Therefore, we instead determine the photometric uncertainties separately using a noise model fit
to an ensemble of $\sim2000$ nearby stars (see Figure~\ref{fig:rms_model}). 
This process is complicated by two factors.
First, the flux from SN 2014J changes significantly in time. Second, SN 2014J
resides in front of a bright galaxy and thus ``sees'' a higher effective sky level
than other sources in the area. 

Our photometric uncertainty model includes Poisson noise contributions from both
the source and the sky, plus an additive term due to unknown systematic errors.
Mathetmatically, the RMS deviation (RMSD) is:
\[
\rms(\, \estar \, ,\,nPix \, , \, \eskpp) =
\sigma_{r} + 
\frac{\sqrt{\estar + nPix \cdot \eskpp}}{\estar}
\]
where $\estar$ is the total number of electrons in the star, nPix is the 
number of pixels in the photometric aperture, $\eskpp$ is the number of sky
electrons per pixel, and $\sigma_r$ is a constant noise floor due to unknown systematics. Though simplistic, this model provides a good description
of the observed photometric uncertainty. 
%{\bf was there a science reason to express this in electrons versus counts, or is it just mathematically simpler?}

Due to the large pixel scale of KELT-N, sky level is a major contributor to the
photometric uncertainty and the dominant contribution for sources fainter than
$V \approx 10.5$. To capture the dependence on sky level, we create three (equal-N) bins of per-pixel sky counts and separately measure RMSD within each bin. We then fit the uncertainty model above to all three bins simultaneously
to determine nPix and $\sigma_r$ (see Fig.~\ref{fig:rms_model}). For the east, we find nPix = 11.94 and $\sigma_r$ = 0.00644. For the west, we find nPix = 7.98 and $\sigma_r$ = 0.00860.
%{\bf I believe this is referring to the different sky ranges in Fig 1?  If so we should point to the figure again.}

Based on reference image measurements, we adopt a a sky excess of 1360 counts for both eastern and western data. Combined with the nPix and $\sigma_r$ values fit
above, we are able to compute robust photometric uncertainties for each data point
in our final SN 2014J light curves.

\subsection{Conversion of instrumental to physical flux units}

\begin{figure}
\centering
\includegraphics[width=\textwidth]{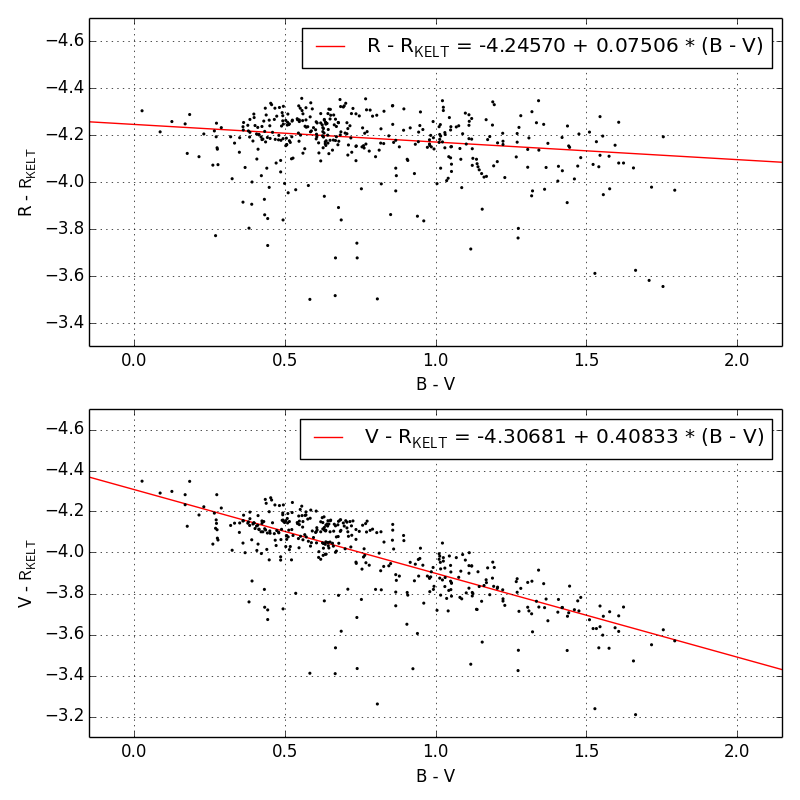}
\caption{Relationships between the KELT-North instrumental system and Johnson 
$V$ and $R$ magnitudes (see the text).
\label{fig:offset}}
\end{figure}

We empirically determine the relationship between KELT instrument response and 
physical flux in two steps. First, we combine KELT star fluxes with catalog data 
from Tycho-2 \citep{2000A&A...355L..27H} and UCAC4 \citep{2013AJ....145...44Z} 
to measure the offset between the KELT instrumental system and standard filters. 
We then use this offset to directly relate the KELT count rate to a flux-calibrated 
star in the Johnson system. We identified stars common to both KELT (east and west) 
object lists and cross-matched these with the Tycho-2 and UCAC4 catalogs. In total, 
we found 745 KELT sources with Tycho-2 and UCAC4 entries. We then robustly fit a
straight line using the Theil-Sen estimator \citep{Sen1968}
to $(R - R_K)$ vs. $(B - V)$ to 
determine the KELT instrumental offset (Figure \ref{fig:offset}).

%}

%{\bf
%The reference star's flux at this apparent magnitude is
%[insert flux here = $ZP \times 10^{-mag / 2.5}$]
%erg s$^{-1}$ cm$^{-2}$ \AA$^{-1}$
%and produced a count rate in the KELT system of
%[insert counts/sec].

We assume for simplicity that SN 2014J has the color of an A0V star.
An A0V star at 0th mag has $R$-band flux of $1.75 \times 10^{-9}$ erg s$^{-1}$ cm$^{-2}$ \AA$^{-1}$ \citep{allens}
and produces a count rate in the KELT system of 
$\sim 3.328 \times 10^{9}$ ADU s$^{-1}$.
%The effective width of the KELT passband, 
%which is similar to but somewhat broader than an $R$ passband, is $\sim 3180$ \AA.
Multiplying by the effective width of the KELT bandpass ($\approx 3180$ \AA), we obtain
an integrated A0V star $R$-band flux of $\sim 5.565 \times 10^{-6}$ erg s$^{-1}$ cm$^{-2}$.
Thus a count rate of 1 ADU s$^{-1}$ corresponds to a total flux in the KELT system of 
$1.672 \times 10^{-15}$ erg s$^{-1}$ cm$^{-2}$.
%[insert total flux from above] / [insert total count rate].
%}

%{\bf 
Finally, we adopt a nominal distance to M82 of $d=3.5$ Mpc \citep{dalcanton09}. Thus, we obtain a final empirical relation between the KELT observed count rate and the total emitted flux at the source of
%[insert total flux from prev paragraph] 
$1.672 \times 10^{-15}$ erg s$^{-1}$ cm$^{-2}$
$\times 4\pi d^2$
erg s$^{-1}$ = $2.451 \times 10^{36}$ erg s$^{-1}$
= 1 ADU s$^{-1}$.

In summary, 1 ADU s$^{-1}$ in the KELT system corresponds to $2.451 \times 10^{36}$ erg s$^{-1}$ at the source (not including any effects of extinction). 
%{\bf FIXME, wordy!}
%}

%\subsection{Empirical Light Curve Representation}
%{\bf [Rob to write, describing the empirical Fourier representation.]}

\section{Results\label{sec:results}}

The full KELT-N light curve of SN 2014J is shown in Figure~\ref{fig:peak_time}. To our
knowledge, this is the most complete, high cadence light curve of this SN 
spanning the entire event yet reported. In this section, we report the results
of analyzing the features of the light curve, specifically the time of initial explosion, short-timescale
variability, the peak brightness time, the secondary bump, and the late-time
plateau.

\subsection{Time of initial explosion\label{sec:t0}}
The KELT-N light curve covering the explosion time of SN\,2014J, along with the intermediate Palomar Transient Factory (iPTF) narrow-band data is
analysed in an accompanying paper \citep{2014arXiv1410.1363G}. The 
extrapolation needed to determine the onset of the optical light from
the supernova is model-dependent, since it takes into account the possibility
that the early emission has contributions besides the radioactive decay of
$^{56}$Ni, e.g, the effect of shock-heated material of the progenitor,
a donor star or the circumstellar medium. Furthermore, radioactivity
arising in the outer parts of the exploding start could produce a
different signature and light curve function to be fitted to the data
to obtain $T_0$. When all the considered alternatives are included, we 
concur with the best fit of \citet{2014ApJ...783L..24Z}: Jan 14.75 UT,
with a systematic uncertainty of $\pm 0.3$~d, due to model dependence. 
We refer to \citet{2014arXiv1410.1363G} for details.

\subsection{Time of maximum light and total rise time}
\begin{figure}
\centering
\includegraphics[width=\textwidth]{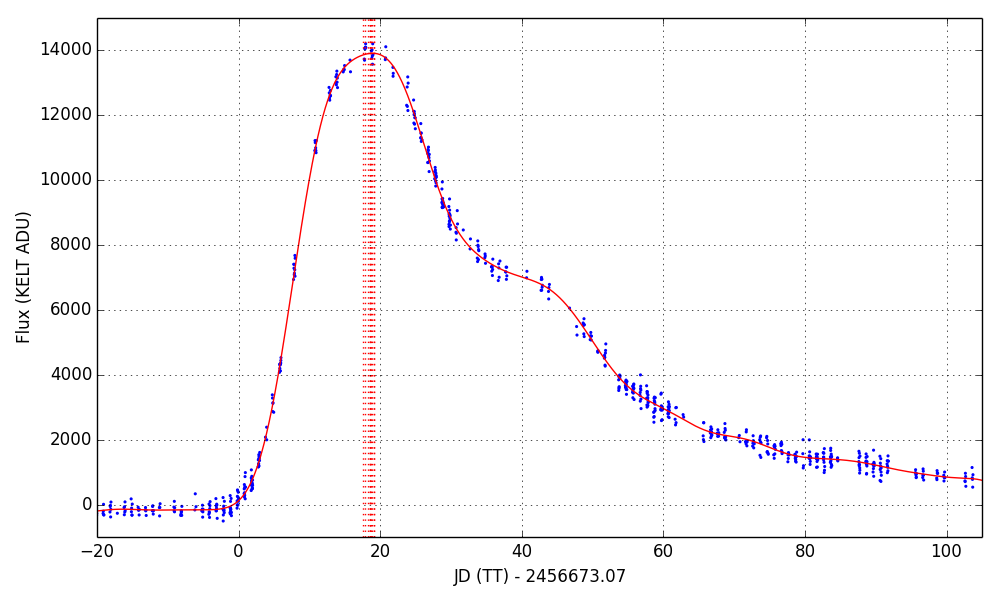}
\caption{KELT light curve of SN2014J (western data). Also shown is the
range of peak time values (vertical lines) determined from best-fit 
empirical Fourier models (solid curve; see the text).
%{\bf [Rob to replace figure.]}
\label{fig:peak_time}}
\end{figure}

%{\bf [Rob to write, focus on the time of peak max.]}
We investigate the time of maximum light by modeling the 
%light curve empirically, modeling the 
KELT-N light curve empirically using
Fourier series and then extracting the time 
%of maximum light
at the peak of the Fourier model
(Figure~\ref{fig:peak_time}).
%We compare this value to the predicted time of maximum in synthetic SN light
%curves based on the known KELT-North instrument response. 
We emphasize that this Fourier representation is not physical and does not
map on to physical parameters as are often used in detailed SN light curve
models. Our intent is to characterize this important property of the light curve,
the peak time, in terms of pure light curve shape parameters.

%We use Fourier series of varying harmonic content to empirically model the
%SN 2014J light curve. We use these models to conveniently characterize the
%temporal light curve appearance. 
%In particular, we measure the time of peak brightness and the delay of the secondary hump and also characterize short-timescale flux variations as a function of time.

%\newcommand{\fmodarg}{\frac{2 \pi nt}{\textrm{T}}}
\newcommand{\fmodarg}{2 \pi nt \, / \, \textrm{T}}
Our Fourier model is a combination of linear polynomial and Fourier terms. The
polynomial is effectively a boundary condition, required because the initial and final fluxes are not equal. The Fourier series representation is:
\[
F(t, \, N) = 
a + bt
+
\sum\limits_{n=1}^N {c_i \sin(\fmodarg) + d_i \cos(\fmodarg)}
\]
where T is the duration of the modeled segment.

We performed this Fourier fitting to the western data using a variety of 
Fourier terms, ranging from
5--14, and we adopt the spread of model light curve maxima 
(represented by vertical lines in Fig.~\ref{fig:peak_time})
as indicative of the systematic uncertainty in the peak time.
Using the ensemble of Fourier series fits to the data, we find that peak flux occurs at 2456691.12 $\pm 0.48$ (JD$_{\rm TT}$).
%2456691.1180253718 $\pm 0.48$ (JD$_{\rm TT}$).

%We separately predict the time of maximum light using synthetic supernova light
%curves generated using the KELT-North instrument response curve.
%{\bf This method is still under investigation ...}

%{\bf [Rob to do a second time adding in the east data to see if that improves 
%peak timing precision.]}

Putting together our updated estimate of the initial explosion time with the
updated estimate of the peak time, we can obtain an estimate of the total
combined rise time for SN2014J in the KELT-N filter to be $18.6 \pm 0.6$~d.

%%Models suggest rise time is 18.92 days (peak - start).

\subsection{Secondary bump}
\begin{figure}
\centering
\includegraphics[width=\textwidth]{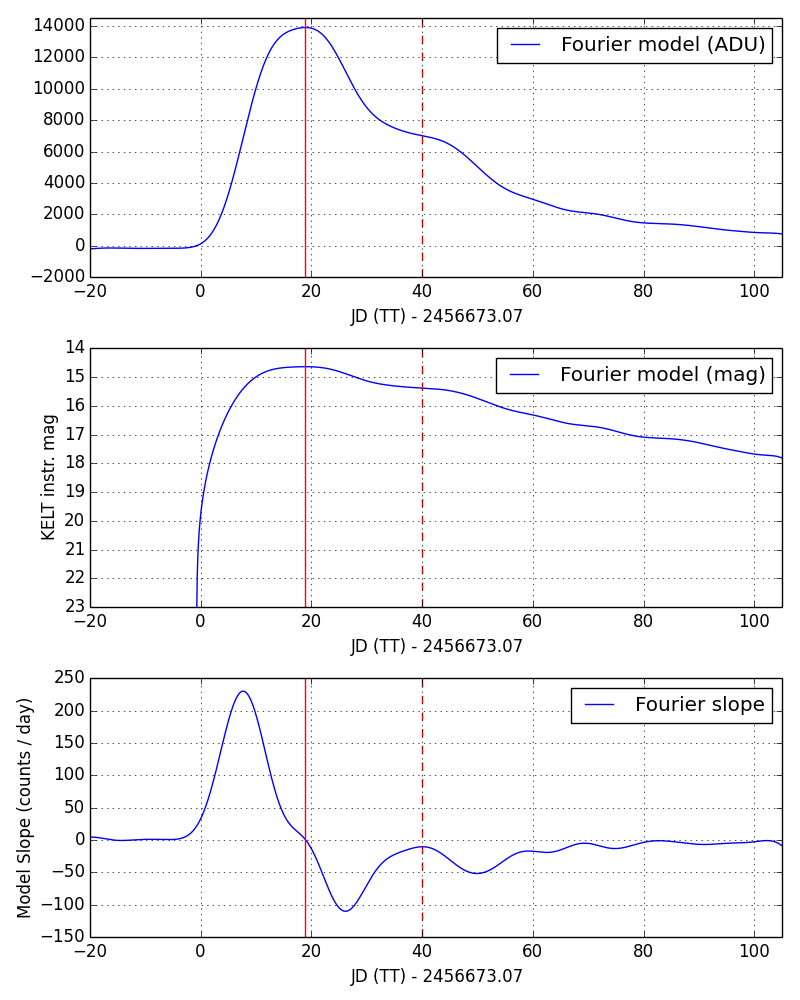}
\caption{Fourier representation of the light curve used to quantify
the time of the secondary bump (see the text).
(Top) Fourier representation of the light curve in flux units. 
(Middle) Same as top but in magnitudes. 
(Bottom) Slope in flux per unit time.
In each panel, the time of primary peak is represented by the solid
vertical line (corresponding to the peak in the top and middle panels), 
and the time of secondary bump is represented by the
dashed vertical line (corresponding to the peak in the bottom panel).
%{\bf [Rob to replace figure.]}
\label{fig:peakhump}}
\end{figure}

We observe a secondary ``bump" in the SN\,2014J light curve approximately
40 days following the initial explosion and approximately 20 days after
the peak brightness. To objectively quantify the precise time of the 
secondary bump, we utilized our empirical Fourier representation of the
light curve to locate the time when the model slope is closest to zero.
Figure~\ref{fig:peakhump} illustrates this approach graphically.
From this analysis, we locate the time of the secondary bump at
%40.08 days
%{\bf xx}
%following the initial explosion and 
%{\bf yyyy} 
21.16~d after the first peak.

\begin{figure}
\centering
\includegraphics[width=\textwidth]{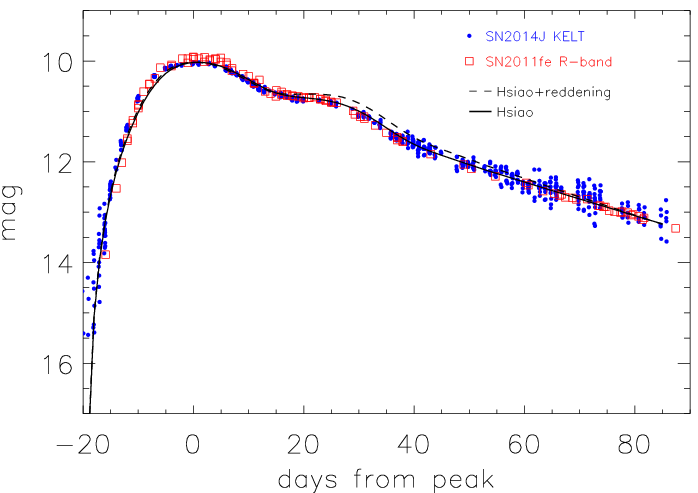}
\caption{The KELT-N light curve of SN\,2014J (blue circles) is shown with the $R$-band data of SN\,2011fe from \citep{2013NewA...20...30M} (red squares) and the
synthetic model photometry (solid line) based on the empirical SNIa SED of \citet{2007ApJ...663.1187H}. The dashed line shows the expectation
from the model SED once the effect of reddening is taken into account, i.e., including the effect of change of the effective wavelength. Note that the
magnitudes from SN\,2011fe are plotted exactly as tabulated in \citep{2013NewA...20...30M}, i.e., without any adjustment for "stretch" or in the vertical axis.
The synthetic models based on \citet{2007ApJ...663.1187H} have been shifted to match the peak brightness of SN\,2014J.}
\label{fig:sn11fe}
\end{figure}

Such secondary bumps are a common feature of SNe at near-IR wavelengths and
the red end of the optical spectrum. Figure~\ref{fig:sn11fe}
shows the KELT-N photometry along with the $R$-band data of SN\,2011fe
from \citet{2013NewA...20...30M} and the synthetic model photometry based on
the empirical SNIa SED of \citet{2007ApJ...663.1187H}, where we show
both the predictions from the integration of the SED over the KELT-N
transmission function with and without the additional effect of reddening
in M82 and in the Milky-Way. For the latter we assume a reddening 
wavelength dependece as parameterized in \citet{1999PASP..111...63F}, 
using the best-fit extintion parameters in \citet{2014ApJ...788L..21A}.

The good match to the template lightcurve reinforces the conclusion 
that SN\,2014J belongs to the class of core normal SNe~Ia, although we
note that the match is better when the reddening correction is {\em not}
included, especially around the secondary maximum, a somewhat unexpected result.
The differences may reflect a possible inaccuracy in the KELT system transmission curve, the calibration using stellar colors, or simply 
intrinsic differences among SNe~Ia.

\subsection{Short-timescale light variations}
%{\bf [Rob to write, focus on the ``smoothness" of the lightcurve at the time scales of minutes/hours never before probed. There could be evidence for/limits on short-lived isotopes or anisotropies in the explosion that have never been sought for.]}
Because of the fortuitously high cadence of observations achieved for the SN 2014J
observations from pre- to post-explosion, the KELT-N light curve affords the 
opportunity to explore the degree of ``smoothness" of the light curve as a function
of time. This in turn can provide evidence for, or limits on, short-lived isotopes
or inhomogeneities in the explosion or the medium 
surrounding it.

We characterize the intrinsic short-timescale variations of the 
light curve by measuring the r.m.s.\ of the light curve in several ways.
First, we model the light curve as a whole using a high-order
Fourier series representation, and measure the r.m.s.\ scatter relative
to this fit. Note that the Fourier representation is not physical, rather
it is a convenient way to empirically represent the overall light curve
and isolate the short-timescale variations. In this way, we measure an
overall r.m.s.\ of $5.1\%$. Near peak SN brightness, the r.m.s.\ scatter
decreases to $1.49\%$. This r.m.s.\ is very nearly equal to the 
expected instrumental precision.

\begin{figure}
\centering
\includegraphics[width=\textwidth]{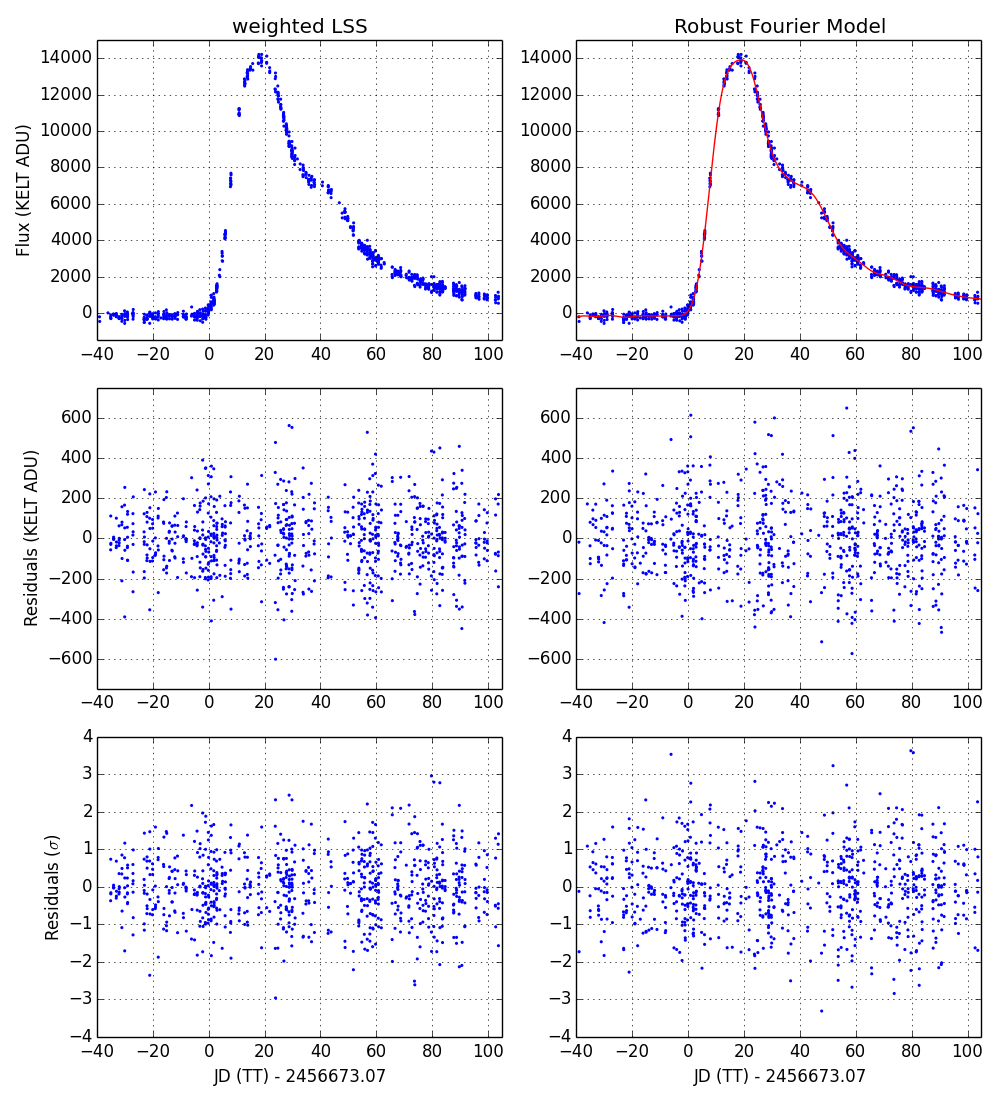}
\caption{Comparison of light curve intra-night scatter using two different fitting methods. Within each night with three or more data points, we fit a straight line to all data points. The top row depicts the SN2014J (west) light curve. The middle row shows the raw residuals in ADU (data points with the straight line fit subtracted). In the bottom row, residuals are divided by the uncertainties. Each column uses a different fitting method. Left uses weighted least-squares and right uses scatter about the empirical Fourier model fit. See the text.
\label{fig:residuals}}
\end{figure}

In addition, we have measured the r.m.s.\ variations on a night-to-night
basis in order to explore whether there may be changes in the 
%intrinsic 
%{\bf what do we mean by the word "intrinsic" here?}
short timescale variations as the supernova progressed. To do this, on
each night possessing at least 3 measurements, we fit a linear trend
and then measure the r.m.s.\ of the measurements on that night relative
to the trend line. Figure~\ref{fig:residuals} shows these nightly r.m.s.\ measurements as
a function of time. Based on these data, we do not observe statistically
significant short-timescale variations on most nights, and therefore we
do not observe statistically significant trends in the short timescale
variations over time. The data do permit us to place an upper limit on the
short timescale variations of 
4.47\% (3$\sigma$) in the time near and shortly following peak brightness.

In Fig.~\ref{fig:residuals}
we represent these r.m.s.\ variations on a 
night-by-night basis. In addition, since as mentioned above these variations are not significantly larger than the instrumental limit, we can use the empirical conversion beween flux in the KELT system to emitted flux at the source (see Sec.~2.6) to represent these measured 
nightly variability limits as limits on the amount of intrinsic variation at the source in erg s$^{-1}$.
%{\bf [Rob to add a Figure with nightly variations and also converting those into erg/s units.]}
On this basis, we can set a 3$\sigma$ limit of 
$3.34\times10^{36}$
erg~s$^{-1}$ for the variations at the source at peak brightness,
and similarly a 3$\sigma$ limit of 
$8.7\times10^{36}$
erg~s$^{-1}$ for the variations over the entire set of KELT observations.
%{\bf [Ariel, say something about this limit.]}

\section{Conclusions\label{sec:conclusions}}
We have reported a complete light curve of the bright M82 SN\,2014J observed
with high photometric cadence from before the explosion, through the early rise and peak light,
and through the secondary bump and beyond to $\sim$100~d past peak brightness.
The KELT light curve confirms that SN\,2014J is a nearby replica of the SNe~Ia
used for precision distance estimates in cosmology. In \citet{2014arXiv1410.1363G}
we examined the first hours after the explosion to conclude that there is evidence
for sources of luminosity in the very early light curve that would indicate
either shock-heating of the SN ejecta, interaction with circumstellar matter or
a companion star, or the presence of radiactive elements near the surface of the exploding star.

In this work we have extended the study to the entire KELT dataset.
We have, for the first time, performed a study of the temporal evolution of a SNIa
that includes the very short timescales of just a few minutes, corresponding
to physical length scales $\lsim{10 R_{\odot}}$ of the expanding SN ejecta. We find 
that any perturbation to the diffuse light emission is smaller than $8.7\times10^{36}$
erg~s$^{-1}$ for the variations over the entire KELT dataset, starting well before the 
explosion. The implications, both with regard to the potential presence
of short-lived radioactive material near the surface of the progenitor, and/or the
small scale structure of the circumstellar medium, will have to be explored with
detailed modeling, currently not available. Thus, the KELT dataset provides new
observational opportunities for the theoretical understanding of Type Ia supernovae.

\acknowledgments
We thank Karen Collins for useful pipeline and data processing discussion. We acknowledge support for the KELT project through the Vanderbilt Initiative in Data-intensive Astrophysics, Ohio State University, and Lehigh University.  Construction of KELT-N was supported by the National Aeronautics and Space Administration under grant No. NNG04GO70G issued through the Origins of Solar Systems program.  AG acknowledges stimulating discussions with Claes Fransson and Markus Kromer, as well as financial support from the Swedish Research Council.

%\begin{figure}
%\centering
%\includegraphics[width=\textwidth]{figures/KN26_colors_M82.png}
%\caption{KELT + UCAC4. Robust straight-line fits are plotted in red.}
%\label{fig:my_label}
%\end{figure}

%\begin{figure}
%\centering
%%\includegraphics[width=\textwidth]{figures/SN2014J_full_west.png}
%\includegraphics[width=\textwidth]{figures/SN2014J_west_full_new.png}
%\caption{KELT light curve of SN2014J (western data).
%{\bf [Rob to replace figure.]}
%\label{fig:west_flux_all}}
%\end{figure}

%\begin{figure}
%\centering
%\includegraphics[width=\textwidth]{figures/EW_mismatch_2.png}
%\caption{Disagreement between KELT eastern and western light curves.}
%\label{fig:ew_mismatch}
%\end{figure}

%\begin{figure}
%\centering
%\includegraphics[width=\textwidth]{figures/SN_model_comparison.png}
%\caption{SN models do not fit.}
%\label{fig:model_comparison}
%\end{figure}

%\begin{figure}
%\centering
%\includegraphics[width=\textwidth]{figures/plateau_white_west.png}
%\caption{KELT data suggest the existence of a late-time flux plateau. Flux does not return to the pre-explosion zero point at late times. Further, KELT-observed flux begins to exceed model predicitions shortly after the second hump.}
%\label{fig:my_label}
%\end{figure}

%\begin{figure}
%\centering
%\includegraphics[width=\textwidth]{figures/late_time_slope_line.png}
%\caption{KELT data suggest the existence of a late-time flux plateau. Flux does not return to the pre-explosion zero point at late times. A simple linear fit to the data up through day 105 indicates that the late-time data points have not continued the same steady decline but instead have reached a plateaus state.}
%\label{fig:slope_line}
%\end{figure}

\end{document}